\documentclass[twocolumn,showpacs,preprintnumbers,amsmath,amssymb]{revtex4}

\usepackage{graphicx}
\usepackage{dcolumn}
\usepackage{bm}

\begin{document}
\title{\bf Unified quaternionic description of charge and spin transport and intrinsic non-linearity of spin currents }
\author{T.~P.~Pareek}
\affiliation{Harish Chandra Research Institute \\ Chhatnag Road, Jhusi,
Allahabad - 211019, India
}%
\begin{abstract}
We present a unified theory of charge and spin transport using quaternionic formalism.
It is shown that both charge current and spin currents can be combined together to form
a quaternionic current. The scalar and vector part of quaternionic currents correspond to charge and spin current respectively. We formulate a unitarity condition on the scattering matrix for quaternionic current conservation. It is shown that in presence of
spin flip interactions a weaker quaternionic unitarity condition implying charge
flux conservation but spin flux non conservation is valid. Using this unified theory we find that spin currents are intrinsically non linear. Its implication for recent experimental
observation of spin generation far away from the boundaries are discussed.
\end{abstract}
\maketitle

During the past decade the study of Spin transport has emerged as an important area
of research in condensed matter.One of the important questions in spintronics is generation and manipulation of
spin currents in semiconductors by means of electric field and spin-orbit coupling.    Experimental observation of spin-Hall effect, both optically\cite{Kato,Sih} and electronically\cite{Tinkham},
acted as a catalyst for recent theoretical works \cite{Halperin,Datta,Pershin}. Theoretically, the spin transport has been addressed
using a variety of methods, e.g., drift-diffusion\cite{Halperin,DasSarma},Kubo formalism\cite{Sinova}, and the scattering theory of transport\cite{pareek,scheid,land-but,Aharony}. The scattering theoretical formulation is particular
suited for the mesoscopic system  since it takes into account both quantum coherence effects and effects of measurement ({\it i.e.} of leads) into account. A detailed account of scattering theory for mesoscopic charge transport has been available in Ref.\cite{Buttiker-prb46}. 
The essential
ingredients of scattering theory are formulated in terms of scattering matrix\cite{Buttiker-prb46}. For
charge transport the charge conservation (global U(1) gauge symmetry) constraints demands that scattering matrix be unitary, {\it i.e.}, $\bf{S}\bf{S}^{\dagger}=\bf{S}^{\dagger}\bf{S}=\bf{I}$.

The situation for spin transport is rather subtle (see Ref.\cite{Frohlich}). In presence of
spin flip interactions (Spin-Orbit interaction, magnetic impurities etc.) though it is generally
stated that spin currents are not conserved however a mathematical equation describing spin
flux non conservation in terms of scattering matrix is still lacking. In the
absence of such a guiding equation it is not clear which specific combination of the
elements of scattering matrix should occur in the calculation of spin currents as opposed to charge current where unitarity of scattering matrix comes to rescue. This has lead to
multitude of equation for spin currents in the literature (see Ref.\cite{pareek,scheid,land-but,Aharony}).

On the other hand it is
well know that scattering matrix of a SO coupled system is {\it quaternionic}\cite{Halperin1, Bennaker} reflecting the break down of spin rotational symmetry. A quaternion $\bm{\hat{q}}$ is defined as  $\bm{\hat{q}}=q^{0}\otimes \bm{I}_{2}+i\bm{q}\cdot\bm{\sigma}$ with $(q^{0},\bm{q})\in \mathcal{C}$, where $\bm{q}$ is a vector and
$\bm{\sigma}$ is a vector of Pauli matrices. The essential features of quaternion is that it combines the scalar and vector quantities into a single entity. The quaternion $\bm{\hat{q}}$ can be viewed as having scalar and vector part($\bm{\hat{q}}\equiv[Scalar,Vector]$).
Intuitively it
is also clear that the charge currents are scalar in nature while spin currents are of vectorial nature.
Therefore, it would be natural to look for a unified description of
charge and spin currents in terms of {\it quaternionic currents}. In the present article we show that
both charge and spin currents can be combined together in terms of a
{\it quaternionic current}, $\mathcal{I}=[I^{q},\bm{I^{s}}]$, where scalar part
gives usual particle(charge) current while the vector part corresponds to the
spin current. We further show that this naturally
introduces a {\it quaternionic metric} on the Hilbert space. We formulate a Strong Unitarity 
condition for the quaternionic scattering matrix implying conservation of {\it quaternionic metric} and show that it is equivalent to conservation of both charge and spin currents. Further it is shown that in presence of a spin flip interaction a weaker form of
quaternionic unitarity condition is valid-implying particle flux conservation but spin flux
non conservation. Using this weaker unitarity condition we derive equation for spin currents
in two terminal system and study its implications.

Before we proceed further it will be helpful to look at charge and spin currents separately.
To elucidate the essential physics we confine our discussion to the case of two terminals systems. The Spin orbit coupled sample is of finite extent and is connected to reservoirs via the ideal leads. We assume that the leads have negligible spin-orbit coupling and spin relaxation so that spin currents are well defined in the leads far away from the sample-lead boundary.  To describe the spin resolves channels in the lead we can choose
a global spin quantization axis in all the leads along $\hat{z}$ direction ( spin resolved wavefunctions are
eigen function of $\sigma_{\hat{z}}$).
The incoming and outgoing spin resolved current amplitudes in leads is given by
column vector $\bm{A}$ and $\bm{B}$ where 
$\bm{A}$=$[\bm{A}_{L}, \bm{A}_{R}]^{T}$ (here $T$ means transpose) and similar defining equation holds for vector $\bm{B}$. Here $\bm{A}_{\alpha}$=$(a_{\alpha,1}^{\sigma},a_{\alpha,1}^{-\sigma}......
a_{\alpha,N_{\alpha}}^{ \sigma},a_{\alpha,N_{\alpha}}^{-\sigma})^T$($\alpha$=$L$ or $R$ corresponding to left and right lead where $a_{\alpha\,m}^{\sigma}$ is annihilation operator in channel $ m\,\sigma$) is a column vector with $N_{\alpha}^{\pm\sigma}$ components
corresponding to number of spin up and down channels. The $\bm{B}$ and $\bm{A}$ are related to each other via $K\times K$ scattering matrix ($K=2N_{L}+2N_{R}$)

\begin{equation}
\bm{B}=\bm{S}\bm{A}
\label{def_sc1}
\end{equation}

{\it Charge current or Particle Flux}: The incoming and outgoing charge currents are given by the the bilinear combinations $\bm{A^{\dagger}A}$ and $\bm{B^{\dagger}B}$. Mathematically these bilinear combinations are in-fact inner product with respect to {\it Identity metric}, $\bm{\Lambda}$ given by,
\begin{equation} \bm{\Lambda}=[\{\bm{I}_{2}\otimes\bm{I}_{L},\bm{0}\},\{\bm{0},\bm{I}_{2}\otimes\bm{I}_{R}\}]
\equiv[\{\bm{\Lambda}_{L},\bm{0}\},\{\bm{0},\bm{\Lambda}_{R}\}]
\label{identity-metric}
\end{equation}
where $\bm{I}_{2}$ being two by two identity matrix in spin space and $\bm{I}_{L}$, $\bm{I}_{R}$ are $N_{L}\times N_{L}$ and  $N_{R}\times N_{R}$ identity matrix for the left and right lead respectively(corresponding to spatial channel index). In terms of $\bm{\Lambda}$, total incoming and outgoing particle(charge) fluxes can be written as,
\begin{eqnarray}
I^{q}_{in}&=&I_{in,L}^{q}+I_{in,R}^{q} =\bm{A}^{\dagger}\bm{\Lambda}\bm{A}=
\sum_{\alpha=L}^{R}\bm{A}_{\alpha}^{\dagger}
\bm{\Lambda}_{\alpha}\bm{A}_{\alpha} \nonumber \\
&=&\sum_{\alpha=L \,n=1}^{R,N_{\alpha}}(a^{\dagger \,\sigma}_{\alpha\,n}a_{\alpha\,n}^{\sigma}+a^{\dagger \,-\sigma}_{\alpha\,n}a_{\alpha\,n}^{-\sigma})
\label{in_pflux}
\end{eqnarray}

\begin{eqnarray}
I^{q}_{out}&=& I_{out,L}^{q}+I_{out,R}^{q}= \bm{B}^{\dagger}\bm{\Lambda}\bm{B}=\sum_{\alpha=L}^{R}\bm{B}_{\alpha}^{\dagger}
\bm{\Lambda}_{\alpha}\bm{B}_{\alpha} \nonumber \\
&\equiv&\sum_{\alpha=L \,n=1}^{R,N_{\alpha}} (b^{\dagger \,\sigma}_{\alpha\,n}b_{\alpha\,n}^{\sigma}+b^{\dagger \,-\sigma}_{\alpha\,n}b_{\alpha\,n}^{-\sigma}).
\label{out_pflux}
\end{eqnarray}
Using Eq.(\ref{def_sc1}) into Eq.(\ref{out_pflux}) and comparing it with Eq.(\ref{in_pflux}),
we immediately see that charge flux conservation implies conservation of {\it current metric}
$\bm{\Lambda}$,{\it i.e.},
\begin{equation}
\bm{S}^{\dagger}\bm{\Lambda}\bm{S}=\bm{\Lambda}.
\label{charge_flux_cons}
\end{equation}

In other words charge flux conservation conserves the identity matrix. In mathematical language
it implies that on the Hilbert space the {\it identity metric} is preserved.

{\it Spin Currents or Spin Flux}: In view of the above discussion a natural question would be what is the corresponding metric for the spin currents?
In other words how to combine $\bm{A}^{\dagger}$, $\bm{A}$ and $\bm{B}^{\dagger}$, $\bm{B}$ to obtain incoming and outgoing spin fluxes respectively. Toward this end we define, the {\it global spin current metric} $\bm{\Omega}$ whose matrix form is given by,
\begin{equation}
\bm{\Omega}=[\{\bm{\sigma}\otimes \bm{I}_{L},\bm{0}\},\{\bm{0},\bm{\sigma}\otimes \bm{I}_{R}\}]\equiv [\{\bm{\Omega}_{L},\bm{0}\},\{\bm{0},\bm{\Omega}_{L}\}]
\label{spc_metric}
\end{equation}
where $\bm{\sigma}=(\sigma_{x}\hat{x}+\sigma_{y}\hat{y}+\sigma_{z}\hat{z})$ is vector of Pauli matrices,and $\bm{I}_{\alpha}$ is ($N_{\alpha}\times N_{\alpha}$, $N_{\alpha}$ being number of spin degenerate channels ) identity matrix in lead $\alpha$. In terms of the spin current metric, incoming and outgoing spin fluxes are defined as,

\begin{eqnarray}
\bm{I}^{s}_{in}&=&
\bm{A}^{\dagger}\bm{\Omega}\bm{A} =\sum_{\alpha=L}^{R}\bm{A}_{\alpha}^{\dagger}\bm{\Omega}_{\alpha}\bm{A}_{\alpha} \label{in-spin1} \\
&=&\bm{I}_{in,L}^{s}+\bm{I}_{in,R}^{s}=\left[I^{s,x}_{in},I^{s,y}_{in},I^{s,z}_{in}\right] \label{in-spin2} \\
&=&\sum_{\alpha=L \,n=1}^{R,N_{\alpha}}\left[ (a^{\dagger \,-\sigma}_{\alpha\,n}a_{\alpha\,n}^{\sigma}+a^{\dagger \,\sigma}_{\alpha\,n}a_{\alpha\,n}^{-\sigma}), i(a^{\dagger \,\sigma}_{\alpha\,n}a_{\alpha\,n}^{-\sigma}-a^{\dagger \,-\sigma}_{\alpha\,n}a_{\alpha\,n}^{\sigma}),\nonumber \right. \\
& &\left. (a^{\dagger \,\sigma}_{\alpha\,n}a_{\alpha\,n}^{\sigma}-a^{\dagger \,-\sigma}_{\alpha\,n}a_{\alpha\,n}^{-\sigma}) \right.]
\label{in-spin3}
\end{eqnarray}

\begin{eqnarray}
\bm{I}^{s}_{out}&=&\bm{B}^{\dagger}\bm{\Omega}\bm{B}=
\sum_{\alpha=L}^{R}\bm{B}_{\alpha}^{\dagger}\bm{\Omega}_{\alpha}\bm{B}_{\alpha} \label{out-spin1} \\
&=&\bm{I}_{out,L}^{s}+\bm{I}_{out,R}^{s} =\left[I^{s,x}_{out},I^{s,y}_{out},I^{s,z}_{out}\right],
\label{out-spin2}
\end{eqnarray}
Where the explicit expression for outgoing spin flux can be obtained from Eq.(\ref{in-spin3})
by replacing incoming operator $a$ by corresponding outgoing operators $b$.
From Eq.(\ref{in-spin3}) we notice that {\it longitudinal spin current}(spin current along the spin quantization axis) $I_{in}^{s,z}$ corresponds to the net spin polarization flowing in this is the standard definition of spin current used in the recent works\cite{pareek,scheid,Aharony,land-but}. The spin currents $I_{in}^{s,x}$ and $I_{in}^{s,y}$ are transverse spin currents(transverse to spin quantization axis) and
corresponds to spin flip processes and deep inside the lead (far away from the
sample-lead boundary) where SO coupling is absent such terms would vanish. However near the
sample lead boundaries these will be non zero and reflects the quantum interference phenomena. These spin flip terms are similar to the complex mixing conductance of circuit theory developed by Brataas, Nazarov and Bauer \cite{Bauer}.
Using Eq.(\ref{def_sc1}) into Eq.(\ref{out-spin1}) and comparing it with Eq.(\ref{in-spin1}), we at once see that spin flux conservation implies that the spin current Metric $\bf{\Omega}$ is preserved,{\it i.e.},
\begin{equation}
\bm{\Omega}=\bm{S}^{\dagger}\bm{\Omega}\bm{S}.
\label{spin_flux_cons}
\end{equation}
The above spin flux conservation equation actually a generalized vector unitarity condition. This is physically plausible also since conservation of a vector quantity would imply
both a conservation of its direction as well its magnitude.

{\it Quaternionic Currents}:
Now we show that charge current and spin currents can be expressed by a single equation using
{\it quternionic formalism}. Towards this end we define {\it quternionic current metric} in lead $\alpha$ is defined as, ${\bm{\mathcal{Q}}_{\alpha}}=[\bm{\Lambda}_{\alpha},\bm{\Omega}_{\alpha}]=[\bm{I}_{2}\otimes\bm{I}_{\alpha},
\bm{\sigma}\otimes \bm{I}_{\alpha}]$, and corresponding {\it quaternionic current} as
$\bm{\mathcal{I}_{\alpha}}=[\mathcal{I}^{q}_{\alpha},\bm{\mathcal{I}}_{\alpha}^{s}]$. The {\it quaternionic current} $\bm{\mathcal{I}_{\alpha}}$ and {\it quternionic current metric } ${\bm{\mathcal{Q}}_{\alpha}}$
can be viewed as having a scalar part and vector part corresponding to particle(charge) flux and spin flux respectively. The full quaternionic current metric for two leads system is expressed
in matrix form as,
\begin{equation}
\bm{\mathcal{Q}}=diag[\bm{\mathcal{Q}}_{L},-\bm{\mathcal{Q}}_{L},\bm{\mathcal{Q}}_{R},-\bm{\mathcal{Q}}_{R}]
\end{equation}
By defining a combined vector creation operators as, $\bm{P}^{\dagger}_{\alpha}=(\bm{A}^{\dagger}_{L},\bm{B}^{\dagger}_{L},\bm{A}^{\dagger}_{R},
\bm{B}^{\dagger}_{R})$, we can write the net {\it quternionic current} $\bm{\mathcal{I}}=[{\mathcal{I}}^{q},\bm{\mathcal{I}}^{s}]$,flowing through the system as,
\begin{eqnarray}
\bm{\mathcal{I}}&=&\bm{P}^{\dagger}\mathcal{Q}\bm{P} \equiv \sum_{\alpha}[\bm{A}^{\dagger}_{\alpha}\mathcal{Q}_{\alpha}\bm{A}_{\alpha}-\bm{B}^{\dagger}_{\alpha}\mathcal{Q}_{\alpha}\bm{B}_{\alpha}] \label{quat_curr} \\
&\equiv &\sum_{\alpha}\bm{A}^{\dagger}_{\alpha}[\bm{I}_{2}\otimes\bm{I}_{\alpha},
\bm{\Omega}_{\alpha}]\bm{A}_{\alpha}- \bm{B}^{\dagger}_{\alpha}[\bm{I}_{2}\otimes\bm{I}_{\alpha},
\bm{\Omega}_{\alpha}]\bm{B}_{\alpha}
\label{quaternion_current}
\end{eqnarray}
It is easy to see that Eq.(\ref{quaternion_current}) is nothing but a combined form of Eq.(\ref{out-spin2}) and Eq.(\ref{charge_flux_cons}). The Eq.(\ref{quat_curr}) is similar to the Eq.(1.11) derived by B\"uttiker in Ref.\cite{Buttiker-prb46} for particle fluxes. We have show that it is valid for quaternionic currents as well.
Now we discusses flux conservation for {\it quaternionic current} $\bm{\mathcal{I}}$. Using the
defining relation Eq.(\ref{def_sc1}, in Eq.(\ref{quaternion_current}) we arrive at 
{\it strong form of quaternionic unitarity condition},
\begin{equation}
\bm{\mathcal{Q}}=\bm{S}^{\dagger}\bm{\mathcal{Q}}\bm{S}= [\bm{\Lambda},\bm{\Omega}]
\label{Q-Uni}
\end{equation}
The above equation is nothing but a unified form of Eq.(\ref{charge_flux_cons}) and Eq.(\ref{spin_flux_cons}) corresponding to charge and spin current conservation.
It is straight forward to see that in presence of spin flip interaction  the Eq.(\ref{Q-Uni}) can not be satisfied for the vector part which would imply spin flux conservation as given by Eq.(\ref{spin_flux_cons}). Towards this end we note that an element of scattering matrix can be expressed as a quaternionic

\begin{equation}
\bf{s}_{\alpha\beta}=\bf{s}_{\alpha\beta}^{0}\otimes\bf{I}_{2}+\bf{s}_{\alpha\beta}^{x}\otimes\bf{\sigma}_{x}+\bf{s}_{\alpha\beta}^{y}\otimes\bf{\sigma}_{y}+\bf{s}_{\alpha\beta}^{z}\otimes
\bf{\sigma}_{z}
\label{quaternion_eq}
\end{equation}
Since the spin flux conservation equation((\ref{spin_flux_cons}) implies that Scattering matrix
commutes with the spin current matrix $\bm{\Omega}$. This implies that each element of
scattering matrix as given by Eq.(\ref{quaternion_eq}) commutes with Pauli matrices which obviously is not true.
Therefore a weaker unitarity condition implying charge flux conservation but spin flux non conservation will be valid, {\it i.e.},
\begin{equation}
\bm{S}^{\dagger}\bm{\mathcal{Q}}\bm{S}=[\bm{\Lambda},\bm{X}]\equiv[\bm{\Lambda},\bm{S}^{\dagger}\bm{\Omega}\bm{S}].
\label{weak-Uni}
\end{equation}
The Eq.\ref{quat_curr} together with Eq.\ref{weak-Uni} provides a unified description of
charge and spin transport in presence of spin flip interactions. Using these two equation 
we find following explicit form of charge and spin current operators,
\begin{eqnarray}
\hat{I}^{q}_{\alpha}&=&\frac{e}{h}\int dE  \sum_{\beta\gamma}\bm{A}^{\dagger}_{\beta}(E)\bm{M}_{\beta\gamma}(\alpha,E)
\bm{A}_{\gamma}(E) \nonumber \\
\bm{M}_{\beta\gamma}(\alpha,E)&=&\bm{\Lambda}_{\alpha}\delta_{\alpha\beta}\delta_{\alpha\gamma}-\bm{S}^{\dagger}_{\alpha\beta}(E)\bm{\Lambda}_{\alpha}\bm{S}_{\alpha\gamma}(E).
\label{charge_curr_op}
\end{eqnarray}

\begin{eqnarray}
\hat{I}^{s,i}_{\alpha}=\frac{1}{4\pi}\int dE \sum_{\beta\gamma}\bm{A}^{\dagger}_{\beta}(E)\Gamma_{\beta\gamma}^{i}(\alpha,E)\bm{A}_{\gamma}(E) \\
\Gamma_{\beta\gamma}^{i}=\bm{\Omega}_{\alpha\,i}\delta_{\alpha\beta}\delta_{\alpha\gamma}-\bm{S}^{\dagger}_{\alpha\beta}(E)\bm{\Omega}_{\alpha\,i}\bm{S}_{\alpha\gamma}(E).
\label{spin_curr_op}
\end{eqnarray}
Where $\bm{\Omega}_{\alpha\,i}=\sigma_{i}\otimes\bm{I}_{\alpha}$ is the spin current metric along $i$ spatial direction,where $i=x,y or z$. Note that the charge current operator (Eq.(\ref{charge_curr_op})) is same as derived by B\"uttiker in Ref.\cite{Buttiker-prb46}.

{\it Average Spin Currents}:
To find the average spin current we note that thermal averages of creation $a^{\dagger\sigma}_{\alpha m}$ and annihilation operator $a^{\sigma}_{\alpha m}$ for
reservoir of unpolarized electrons is,
\begin{equation}
\overline{a^{\dagger\sigma}_{\beta m}(E)a^{\sigma^{\prime}}_{\gamma n}(E^{\prime})}=\delta(E-E^{\prime})\delta_{\beta\gamma}\delta_{m n}\delta_{\sigma\sigma^{\prime}}f_{\beta}
\end{equation}
where $f_{\beta}\equiv f_{\beta}(E,\mu_{\beta})$ is the Fermi distribution function with the chemical potential
$\mu_{\beta}$ of reservoir $\beta$\cite{Buttiker-prb46}. 
Performing the average we find that the
average {\it longitudinal spin current $I^{s z}$}(along the spin quantization axis) for two terminal systems is,
\begin{eqnarray}
I^{s z}_{\alpha}&=& \frac{1}{4\pi}\int dE
\left[ (R_{\alpha\alpha}^{\sigma\sigma}+R_{\alpha\alpha}^{\sigma-\sigma}-
R_{\alpha\alpha}^{-\sigma\sigma}-R_{\alpha\alpha}^{-\sigma-\sigma})f_{\alpha}\nonumber \right. \\ 
& + &
\left. (T_{\alpha\beta}^{\sigma\sigma}+T_{\alpha\beta}^{\sigma-\sigma} -T_{\alpha\beta}^{-\sigma\sigma}-T_{\alpha\beta}^{-\sigma-\sigma})f_{\beta} \right.]
\label{spincurrZ}
\end{eqnarray}
with $\alpha=L, R$ and $\beta\neq\alpha$.
In writing down the Eq.(\ref{spincurrZ}) we have not
assumed time reversal symmetry. In case time reversal is a symmetry, $R_{\alpha\alpha}^{\sigma\sigma}=R_{\alpha\alpha}^{-\sigma-\sigma}$ and $ T_{\alpha\beta}^{\sigma\sigma}=T_{\beta\alpha}^{-\sigma-\sigma}$,$T_{\alpha\beta}^{\sigma-\sigma}=T_{\beta\alpha}^{\sigma-\sigma}$ still the Eq.(\ref{spincurrZ}) predicts a non zero spin current.
The important point to note is that even if one insists on using unitarity condition of
particle flux sector(conservation of ${\bm{\Lambda}}$ in Eq.(\ref{weak-Uni})) still the spin currents can not be written down as a difference of Fermi distribution functions. To see this explicitly we rewrite the eq.(\ref{spincurrZ}) using conservation of ${\bm{\Lambda}}$ as,
\begin{equation}
I^{s z}_{\alpha}=\frac{1}{4\pi}\int dE
\left[ 2(R_{\alpha\alpha}^{\sigma-\sigma}-
R_{\alpha\alpha}^{-\sigma\sigma})f_{\alpha}+{\cal{T}}_{\alpha\beta}(f_{\beta}-f_{\alpha})\right],
\label{spcZ}
\end{equation}
where ${\cal{T}}_{\alpha \beta}=(T_{\alpha \beta}^{\sigma\sigma}+T_{\alpha \beta}^{\sigma-\sigma} -T_{\alpha \beta}^{-\sigma\sigma}-T_{\alpha \beta}^{-\sigma-\sigma})$ and $f_{i}\equiv
f_{i}(E,\mu_{i})$ for $i=\alpha$ or $\beta$. From eq.(\ref{spcZ}) we see that net {\it{longitudinal spin currents}} has equilibrium as well nonequilibrium contribution and these can not be separated. Because for the equilibrium part the spin resolved reflection probabilities
contribute while for the non-equilibrium part corresponding transmission probabilities
contribute. Experimentally one measures the net currents(net spin polarization) and not the transmission or reflection probabilities separately.
Therefore spin currents are intrinsically non-linear in an electrical circuit where the driving
force is electrical voltage. This is not surprising because the gauge conjugate to spin currents is non-Abelian SU(2) gauge which consists of terms describing spin-orbit interaction see Ref.\cite{Frohlich, Tokatly}. While the Eq.(\ref{spincurrZ}) is written in terms of U(1) gauge (electrical voltage) conjugate to charge currents. We stress that the non-linearity of
spin currents does not imply breakdown of linear response theory, rather it is non-applicability of linear response with respect to electrical voltage for spin currents.

{\it Transverse Spin Current:} The expression for average transverse spin currents$ $
can not be written down in terms of spin resolved probabilities.
It turns out to be much more complicated,however, to bring out essential physics we express these currents in terms of spin resolved block scattering matrix as,
\begin{equation}
I^{s, x_{+},y_{-}}_{\alpha}=\int dE \sum_{\beta,\sigma^{\prime}}\left[Tr(\bm{S}^{\dagger \sigma \sigma^{\prime}}_{\alpha\beta}\bm{S}^{-\sigma \sigma^{\prime}}_{\alpha\beta}\pm \bm{S}^{\dagger -\sigma \sigma^{\prime}}_{\alpha\beta}\bm{S}^{\sigma \sigma^{\prime}}_{\alpha\beta})f_{\beta}\right].
\label{x-y-spincurr}
\end{equation}
In the above equation $\alpha$ and $\beta$ can take values either $L$ or $R$ corresponding to left and right lead. Similarly the summation over $\sigma^{\prime}$ can take two values $\sigma$ or $-\sigma$. It is obvious from the Eq.(\ref{x-y-spincurr}) that it can not be
expressed as a difference of Fermi distribution function signifying the non-linearity
as discussed above. In Eq.(\ref{x-y-spincurr}) the spin conserved and spin flip scattering matrix elements (reflection and transmission amplitudes) occur together which is a quantum interference effect. On thermal averaging such an effect would vanish deep inside the lead
far away from the sample lead boundary. However near the boundary this will give rise to oscillatory spin currents where evanescent state plays a role\cite{Aharony,Sablikov}. As remarked earlier these currents are similar to the mixing conductance which appear in the circuit theory of
Brataas, Nazarov and Bauer \cite{Bauer}.

As we have seen above that {\it longitudinal spin currents} survives deep inside the lead and
may give rise to spin currents far away from the boundary. In a recent experiments
Sih.{\ et al.}\cite{Sih} observed spin generation away from the edges of a GaAs sample subjected to an
electrical field. A theoretical explanation of such an effect has been worked out in Ref.\cite{Halperin,Datta,Pershin}. In Ref.\cite{Halperin} it was argued that spin generation far away from the boundary is not possible in the linear regime. Though we can not make a
direct comparison with the experimental results of \cite{Sih}, however, the Eq.(\ref{spincurrZ}) for longitudinal spin current above generically show that 
these are intrinsically non linear
with respect to electric field hence can exist far away from the boundary.

Finally let us discusses the possibility of observing these experimentally.
Most promising technique looks to be an optical measurement, especially using
Magneto-optical Kerr effect(MOKE). In the Kerr effect the rotation of the polarization axis of reflected beam is proportional to the spin polarization along the beam direction\cite{Meier}. Therefore  to study all the component of spin current it would be essential to measure Kerr rotation with different angle of incidence (beam direction). Considering that such experiments are routinely
done for spin studies it should not be very difficult to experimentally
test the effects discussed in this paper.

To Conclude we have presented a unified formulation of spin and charge
transport using quaternionic currents. Unitarity condition of quaternionic currents is formulated and spin currents equation are derived. It would be interesting to extend this formulation to study spin shot noise which might have some
nontrivial bearings on semiconductor spintronics.

\end{document}